\documentclass[
    a4paper,
    prb,
    bibnotes,
    twocolumn,
    superscriptaddress,
    groupedaddress,
    floatfix
]{revtex4-1}
\usepackage{graphicx}
\usepackage{multirow}
\usepackage{amssymb,amsmath}
\usepackage{amsmath}

\usepackage{verbatim}
\usepackage[colorlinks,anchorcolor=blue,urlcolor=blue,citecolor=blue]{hyperref}
\usepackage{color}
\newcommand{\sk}{s\vec{k}}

\newcommand{\Vq}{\vec{q}}

\newcommand{\VK}{\vec{k}}
\newcommand{\Vu}{\vec{u}}
\newcommand{\VI}{\vec{I}}

\begin{document}
\title{Nonlocal hydrodynamic phonon transport in two-dimensional materials}
\author{Man-Yu Shang} 
\author{Jing-Tao L\"u}  
\affiliation{School of Physics and Wuhan National High Magnetic Field Center, Huazhong University of Science and Technology, Wuhan 430074, P. R. China}

\begin{abstract}
We study hydrodynamic phonon heat transport in two-dimensional (2D) materials.
Starting from the Peierls-Boltzmann equation with the Callaway model, we
derive a 2D Guyer-Krumhansl-like equation describing non-local hydrodynamic
phonon transport, taking into account the quadratic dispersion of flexural
phonons. In addition to Poiseuille flow, second sound propagation, the
equation predicts heat current vortices and negative nonlocal thermal conductance in 2D
materials, which is common in classical fluid but has not yet been considered in phonon transport.  
Our results also illustrate the universal transport behavior of hydrodynamics, 
independent of the type of quasi-particles and their microscopic interactions.
\end{abstract}
\maketitle

\section{Introduction}
Macroscopic collective behavior emerges from microscopic many-body interactions
between individual degrees of freedom comprising the system. Hydrodynamics is
one of such macroscopic phenomena. It could originate from different kinds of
microscopic interactions in different materials, ranging from classical
gases and liquids,  to crystal solids\cite{gurzhi_hydrodynamic_1968,beck_phonon_1974,de_jong_hydrodynamic_1995,bandurin_negative_2016,crossno_observation_2016,moll_evidence_2016,de_tomas_kinetic_2014,sellitto_two-dimensional_2015,guo_phonon_2015,cepellotti_phonon_2015,lee_hydrodynamic_2015,levitov_electron_2016},
and to cold atomic gases\cite{cao_universal_2011} or hot nuclear
matter\cite{jacak_exploration_2012}.  Although the microscopic inter-particle
interactions are of different nature, the hydrodynamic behaviors are universal.
They can be described by similar hydrodynamic equations. These
equations can normally be derived from the microscopic equations of motion by
considering physical quantities that are conserved during the inter-particle
collisions, i.e., (crystal) momentum,  energy or particle number.

Although hydrodynamic flow in classical gases and liquids is a common process that can be observed in everyday life,
to observe hydrodynamic transport of (quasi-)particles in crystalline solids is
much more difficult. Conservation of crystal momentum is required during the
inter-particle collisions. This needs high quality samples to reduce extrinsic
scatterings with impurities. It also requires that the intrinsic scattering
between quasi-particles to be normal (N-process), which conserves the crystal
momentum, instead of Umklapp (U-process), which does not.
Furthermore, the
hydrodynamic feature is prominent in spatial confined samples like
one-dimensional (1D) or two-dimensional (2D) materials\cite{RevModPhys.90.041002,lirmp1,wang}, which raises further challenges in their fabrication and characterization.

Due to these limitations, studies on the hydrodynamic transport of
quasi-particles in solid state system are scarce.
Recently, experimental and numerical signatures of hydrodynamic electron
\cite{bandurin_negative_2016,moll_evidence_2016,crossno_observation_2016,gooth_electrical_2017,levitov_electron_2016,guo_higher-than-ballistic_2016,falkovich_linking_2016,alekseev_negative_2016,pellegrino_electron_2016,briskot_collision-dominated_2015,narozhny_hydrodynamics_2015,lucas_hydrodynamics_2018}
and
phonon\cite{cepellotti_phonon_2015,lee_hydrodynamic_2015,cepellotti_thermal_2016,cepellotti_transport_2016,lee_hydrodynamic_2017,ding_phonon_2017,sellitto_two-dimensional_2015,PhysRevB.96.134312}
transport in 2D materials have been reported. For electron transport, negative
nonlocal resistance\cite{bandurin_negative_2016}, violation of Wiedemann-Franz
law\cite{crossno_observation_2016} and large negative
magnetoresistance\cite{gooth_electrical_2017} have been experimentally observed and theoretically
explained\cite{levitov_electron_2016,pellegrino_electron_2016,briskot_collision-dominated_2015,alekseev_negative_2016,guo_higher-than-ballistic_2016}.

Considering the universal behaviour of hydrodynamics, we expect similar transport behaviour may exist for other quasi-particles in solid. We focus on phonons here. Poiseuille flow
and the propagation of second sound have been studied in graphene and similar 2D
materials by numerically solving the semi-classical Boltzmann equation with
inputs from density functional theory
calculation\cite{lee_hydrodynamic_2015,cepellotti_phonon_2015}.
It is suggested that, contrary to three-dimensional materials\cite{beck_phonon_1974,Helium,Bismuth,NaF,STO,machida_observation_2018,BismuthT}, hydrodynamic
phonon transport in 2D materials persists over a much larger  temperature range($50\sim 150$ K) in micrometer scale samples. The quadratic dispersion of graphene ZA
acoustic phonon mode is argued to play an important role in widening the
temperature range\cite{lee_hydrodynamic_2015}.

However, unlike electrons, experimental evidence of phonon hydrodynamic
transport in 2D materials has not been observed, despite recent progress in 3D materials\cite{STO,machida_observation_2018}. Theoretical analysis based on
simplified models may help to identify possible experimental signatures of phonon
hydrodynamics. Considering its universal behavior, it is interesting to ask
whether similar effects observed for electrons can be expected for phonons.
Here, we answer this question from the analysis of a Guyer-Krumhansl (G-K)
equation for 2D materials, which we  derive from the Peierls-Boltzmann equation
with the Callaway model. Importantly, we consider both linear and quadratic
acoustic phonon dispersion, which is critical to 2D materials. We extend the multiscale expansion technique \cite{guo_phonon_2015} to include both linear and quadratic phonon modes in 2D materials. This has not been considered before. 
We show that the G-K equation takes a
familiar form, but the transport coefficients differ from normal Debye model, which assumes linear dispersion of acoustic phonon modes. The viscosity coefficients, the second sound velocity become temperature dependent, contrary to the Debye model. Our results will be useful for further theoretical and experimental study of phonon hydrodynamics in 2D materials.

The paper is organized as follows. The Sec.~\ref{sec:2D} we introduce our model and sketch the derivation of the 2D G-K equation. The technique details are presented in Appendix~\ref{sec:appa}. As limiting cases, in Sec.~\ref{sec:2sound}, we show that our result predicts second sound propagating and Poiseuille flow, which have been studied previously by numerical calculations\cite{lee_hydrodynamic_2015,cepellotti_phonon_2015}. In Sec.~\ref{sec:nonlocal}, as our new prediction, we show that negative nonlocal  heat conductance and heat current cortices may appear in ribbons of 2D materials with local heat current injection. Our conclusions are given in Sec.~\ref{sec:conclusion}.

\begin{figure*}
\includegraphics[scale=0.55]{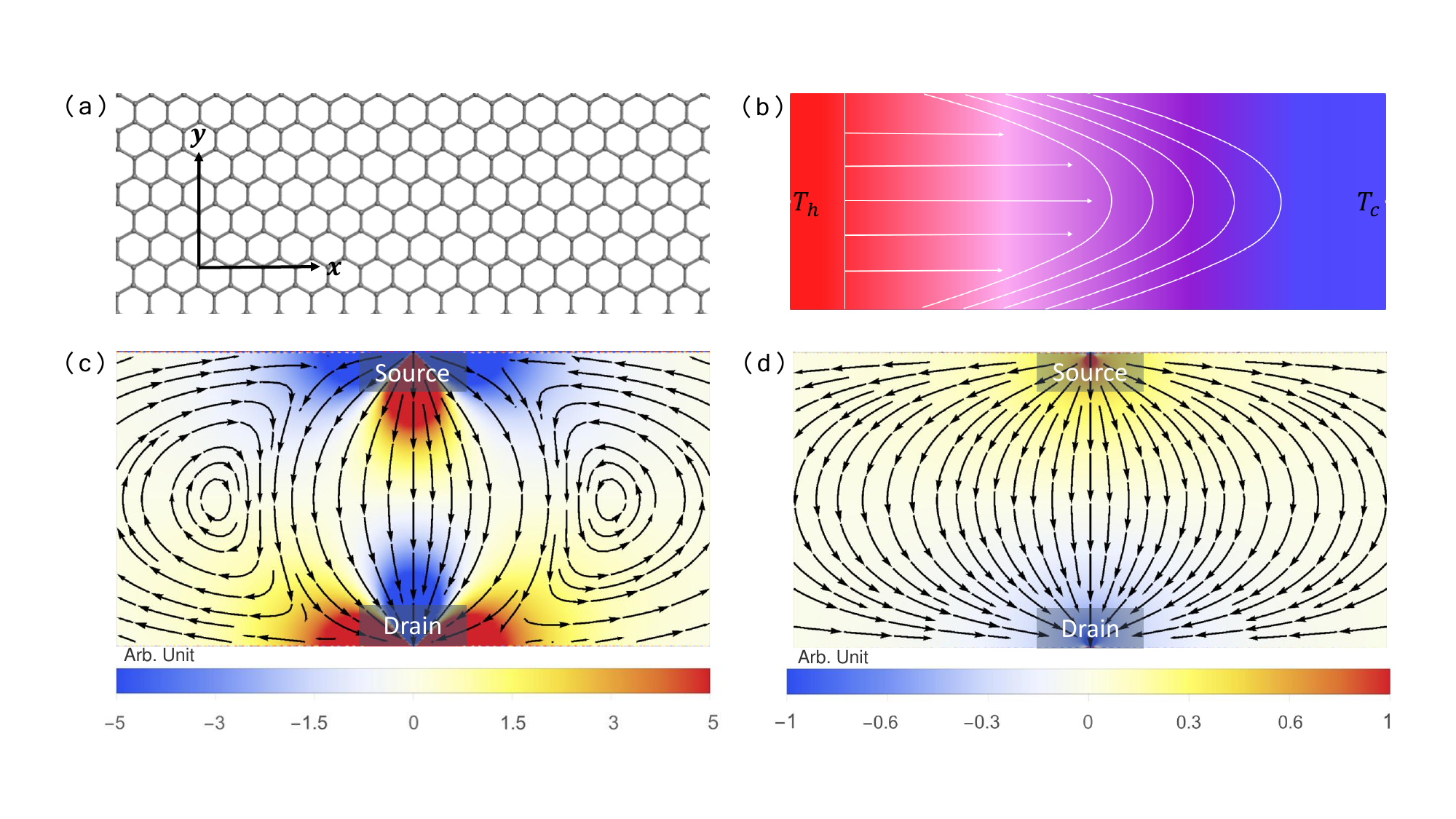}
\caption{(a) A graphene nano-ribbon as a prototype 2D materials showing hydrodynamics phonon transport. (b) Schematic of Poiseuille flow generated by the
temperature difference along the nano-ribbon. The heat flux has a parabolic
distribution across the ribbon. (c) Heat current loops (lines) and temperature distribution (color)
due to the viscosity of the phonon gas as signature of hydrodynamic heat transport. We set $\epsilon=1$.
 (d) The same as (c), but in the diffusive Fourier transport regime with $\epsilon=300$. }
\label{fig:flow}
\end{figure*}

\section{The 2D G-K equation}
\label{sec:2D}
We consider a prototype 2D material. It has one out-of-plane acoustic mode with
a quadratic dispersion $\omega_k = a k^2$ (ZA mode), and two degenerate linear
acoustic modes $\omega_k = v_g k$ (longitudinal and transverse). The magnitude
of the linear group velocity is $v_g$, and the magnitude of the wave vector is
$k=|\vec{k}|$. Here, in the spirit of Debye model, we ignore the possible
anisotropic property and the difference between longitudinal and transverse
branches. We will focus on the effect of ZA mode's quadratic dispersion on the hydrodynamic behavior.

We first sketch the derivation of the 2D G-K
equation\cite{guyer_solution_1966,guyer_thermal_1966}. Our starting point is the
Peierls-Boltzmann equation under Callaway
approximation\cite{callaway_model_1959,allen_improved_2013}
\begin{equation}
\frac{\partial{f_{\sk}}}{\partial{t}}+\vec{v}_{\sk}\cdot\nabla {f}_{\sk}=-\frac{f_{\sk}-f_{R,\sk}^{eq}}{\tau_{R}}-\frac{f_{\sk}-f_{N,\sk}^{eq}}{\tau_{N}}.
\label{eq:pb}
\end{equation}
Here, $s$ is the phonon index, $\tau_N$ is the constant relaxation time for the  N-process, while
$\tau_R$ is that for the resistive scattering process (R-process). It
includes all scattering mechanisms that do not conserve crystal momentum, i.
e., impurity scattering, phonon-scattering and other U-processes.  We take the same $\tau_R$ and
$\tau_N$ for all phonon branches, i.e., taking the wave vector and branch averaged
values.  This is the so-called gray approximation.

The N-process drives the system towards a displaced distribution function
\begin{equation}
f_{N,\sk}^{eq}=\left[e^{\beta_B(\hbar\omega_{\sk}-\hbar\vec{k}\cdot\vec{u})}-1\right]^{-1},
\end{equation}
with $\beta_B=(k_BT)^{-1}$, and $\vec{u}$ is the drift velocity. But the R-process drives the system to an
equilibrium Bose-Einstein distribution
\begin{equation}
f_{R,\sk}^{eq}=\left(e^{\beta_B\hbar\omega_{\sk}}-1\right)^{-1}.
\end{equation}
It has been shown numerically that, within some moderate temperature range ($\sim 100$ K for graphene),
the N-process is orders of magnitude faster than the R-process, meaning
$\tau_R \gg \tau_N$\cite{lee_hydrodynamic_2015,cepellotti_phonon_2015}. When
the system size is much larger than the normal-scattering mean free path $l
\approx v_g \tau_N$.  The relaxation from local to global equilibrium or steady
state is governed by hydrodynamic equations describing the conserved quantities
during the $N$-process, including energy and crystal momentum.

As we know, conservation laws play central roles in the derivation of hydrodynamic equations. We consider energy and crystal momentum conservation here. Both normal (N) and resistive (R) scattering processes conserve total energy, giving
\begin{align}
\sum_s \int \frac{d\VK}{(2\pi)^2} \hbar\omega_{\sk} f_{\sk}
&=\sum_s \int \frac{d\VK}{(2\pi)^2} \hbar\omega_{\sk} f_{N,\sk}^{eq},\label{eq:ec1}\\
\sum_s \int \frac{d\VK}{(2\pi)^2} \hbar\omega_{\sk} f_{\sk}
&=\sum_s \int \frac{d\VK}{(2\pi)^2} \hbar\omega_{\sk} f_{R,\sk}^{eq},\label{eq:ec2}
\end{align}
while only the N-process obeys crystal momentum conservation, giving
\begin{equation}
\sum_s \int \frac{d\VK}{(2\pi)^2} \hbar\VK f_{\sk}
=\sum_s \int \frac{d\VK}{(2\pi)^2}\hbar\VK f_{N,\sk}^{eq}.
\end{equation}

To get the equation governing the dynamics of the conserved quantity, we
multiply by $\hbar\omega_{s\vec{k}}$, integrate over $\VK$ and sum over the phonon index
on both sides of Eq.~(\ref{eq:pb}). We then arrive at an equation describing
energy conservation
\begin{equation}
\frac{\partial{E}}{\partial{t}}+\nabla\cdot\Vq=0,
\label{eq:econsv}
\end{equation}
where
\begin{equation}
E = \sum_s \int  \frac{d\VK}{(2\pi)^2} \hbar\omega_{\sk} f_{\sk}
\end{equation}
is
the total energy density, and
\begin{equation}
\Vq = \sum_s \int \frac{d\VK}{(2\pi)^2} \hbar \omega_{s\VK}\vec{v}_{s\VK} f_{s\VK}
\end{equation}
is the heat current density. The
right hand side of Eq.~(\ref{eq:econsv}) is zero because the scattering
processes conserve energy.

In principle, we can write down similar equation for the crystal momentum
density $\vec{p}$ and its flux from its conservation law during $N$-processes.
The G-K equation can be derived from the resulting momentum balance equation.
This works well for the Debye model with linear phonon dispersion.  But for graphene-like 2D system,
the presence of quadratic dispersion leads to a divergent momentum flux, making
further derivation difficult\cite{phdthesis}. To avoid this, we consider the
heat flux directly, i.e., multiplying $\hbar\omega \vec{v}$ to each term in
Eq.~(\ref{eq:pb}) and summing over all wave vectors and phonon indices. This
leads to
\begin{equation}
\frac{\partial{\Vq}}{\partial{t}}+\frac{\vec{\kappa}}{\tau_R}\cdot\nabla T=-\frac{\Vq}{\tau_R}-\frac{\Vq_1}{\tau_N}.
\label{eq:hd0}
\end{equation}
Here, $\Vq_1$ is the 1st order term of heat flux, which will be explained
below. We have also defined a thermal conductivity tensor as
\begin{equation}
\vec{\kappa}=\tau_R \sum_s \int\frac{d\VK}{(2\pi)^2} \hbar\omega_{\sk}\vec{v}_{\sk}\vec{v}_{\sk}\frac{\partial f_{\sk}}{\partial T}.
\end{equation}

To obtain the hydrodynamic equation, we rely on a multi-scale  expansion
technique developed recently\cite{guo_phonon_2015}.  The expansion is over both space and time as follows:
  \begin{align}
  \frac{\partial{}}{\partial{t}}=&\varepsilon\frac{\partial}{\partial{t_1}}
  +\varepsilon^2\frac{\partial}{\partial{t_2}},\\
  \frac{\partial{}}{\partial{x_{i}}}=&\varepsilon\frac{\partial}{\partial{x_{1i}}}.
  \end{align}
 It is a perturbation expansion over a natural small parameter
\begin{equation}
	\varepsilon = \frac{\tau_N}{\tau_R}.
	\label{}
\end{equation}

That is to say, we consider the situation where the scattering rates of N-process is much larger than that of R-process. We expand the phonon distribution function as follows
  \begin{equation}
  f=f_0+\varepsilon f_1+\varepsilon^2 f_2+\cdots,
  \end{equation}
with $f_0$, $f_1$, $f_2$ the 0th, 1st and 2nd order terms of phonon distribution function. The macroscopic variables can be expressed by the sum of approximate components 
  \begin{align*}
   &E=E_0+\varepsilon E_1+\varepsilon^2 E_2+\cdots,\\
   &E_n=\sum_s \int\hbar\omega_{\sk} f_{n,\sk}\frac{d\VK}{(2\pi)^2}.\\    
  &\Vq=\vec{q}_0+\varepsilon \vec{q}_1+\varepsilon^2 \vec{q}_2+\cdots,\\
  &\vec{q}_n=\sum_s \int\vec{v}_{\sk}\hbar\omega_{\sk} f_{n,\sk}\frac{d\VK}{(2\pi)^2}.\\
  &\vec{\kappa}=\vec{\kappa}_0+\varepsilon \vec{\kappa}_1+\varepsilon^2 \vec{\kappa}_2+\cdots,\\
 &\vec{\kappa}_n=\tau_R \sum_s \int\hbar\omega_{\sk}\vec{v}_{\sk}\vec{v}_{\sk}\frac{\partial f_{n,\sk}}{\partial  T}\frac{d\VK}{(2\pi)^2}.
  \end{align*}

We include the 0th and 1st order terms with $\Vq\approx \Vq_0+\varepsilon
\Vq_1$, $\vec{\kappa} \approx \vec{\kappa}_0+\varepsilon
\vec{\kappa}_1$. According to energy conservation (Eq.~(\ref{eq:ec1})), we know that $E=E_0$, since $E_{n}=0$ for $n>0$.  All these are calculated from the distribution function
$f\approx f_0+\varepsilon f_1$, with $f_0=f_N^{eq}$,
$f_1=f_R^{eq}-f_N^{eq}-\tau_N\left(\partial_{t_1}f_{N}^{eq}+v_{i}\partial_{x_{1i}}{f_N^{eq}}\right)$.
When $\Vu$ is small, we can approximate $f_{N}^{eq}$ as
\begin{equation}
f_{N}^{eq} \approx f_{R}^{eq}+\hbar\beta_B f_{R}^{eq}(f_{R}^{eq}+1)  \VK\cdot\Vu.
	\label{}
\end{equation}

Numerical calculation shows that this is indeed a good
approximation\cite{lee_hydrodynamic_2015}. Taking into account only 
the 0th order term $f_0$, we can get 
 \begin{align}
  E &=\frac{2}{\pi}\frac{(k_BT)^3}{(\hbar v_g)^2}Zeta(3)+\frac{\pi}{24}\frac{(k_BT)^2}{\hbar a} \equiv E_L+E_N,\\
  \Vq_0 &=\frac{3}{\pi}\frac{(k_BT)^3}{(\hbar v_g)^2}Zeta(3)\Vu+
  \frac{\pi}{12}\frac{(k_BT)^2}{\hbar a}\Vu
  \equiv \vec{q}_{0L}+\vec{q}_{0N},\\
  \vec{\kappa}_0 &=\frac{6Zeta(3)}{\pi}\frac{k_B^3T^2}{\hbar^2}\tau_R\VI=C_{L}v_g^2\tau_R\VI.
  \end{align}
  
 Here,$E_L$, $E_{N}$ are the equilibrium energy density of the linear and quadratic
phonon modes, respectively. $\Vq_{0L}$, $\Vq_{0N}$ are the 0th order contribution of the
linear and quadratic modes to the heat flux.  They are both proportional to the
drift velocity $\Vu$. $C_{L} = \left(\partial E_L/\partial T\right) =\frac{6Zeta(3)}{\pi}\frac{k_B^3}{(\hbar v_g)^2}T^2$ is the specific 
heat capacity of the linear modes, $C_{N}$ is defined similarly, and $C=C_{L}+C_{N}$.

 The first-order result of macroscopic variables are 
 \begin{align}
  E_1&=0,\\
  \Vq_1&=0,\\
\vec{\kappa}_1&=\tau_R \frac{\partial \vec{Q}}{\partial T}=\tau_R 	\left(\nabla\cdot\vec{Q}\right)\frac{1}{\nabla T}.
\end{align}
 
The expression of $\vec{Q}$ and calculation details can be found in the Appendix. We arrive at the G-K equation of 2D materials
\begin{equation}
	\frac{\partial{\Vq}}{\partial{t}}+\frac{\kappa_0}{\tau_R} \nabla T+\frac{1}{\tau_R}\Vq=\eta
\big[\nabla^2\Vq+2\nabla(\nabla\cdot\Vq)\big]-
\zeta\nabla(\nabla\cdot\Vq).
\label{eq:gk}
\end{equation}
We have defined the 0th order thermal conductivity as $\kappa_0 = \alpha C_{L}
v_g^2 \tau_R$, the first  and second viscosity coefficients as  $\eta = \beta
v_g^2\tau_N$, $\zeta = \gamma v_g^2\tau_N$, with  $\alpha =1$, $\beta =
\frac{9E_L}{4(3E_L+4E_N)}$, $\gamma = C_{L}/C$, respectively. Equation~(\ref{eq:gk}) with the
above defined coefficients is the central results of this work.  We can see
that, although the form of the G-K equation is the same as the 3D case, the inclusion of
quadratic phonon mode changes its coefficients. Notably, $\eta$ and $\gamma$ become
temperature dependent, while for Debye model with 
three degenerate linear acoustic phonons modes, the coefficients are constant, with $\alpha=1/3$, $\beta=1/5$, $\gamma=1/3$ for 3D and $\alpha=1/2$, $\beta=1/4$, $\gamma=1/2$ for 2D case, respectively.
In the following, we will analyze the consequences of this equation.

\section{Second sound and Poiseuille flow}
\label{sec:2sound}
We now analyze the consequence of Eq.~(\ref{eq:gk}). In this section, we show that, the numerical results in previous studies\cite{lee_hydrodynamic_2015,cepellotti_phonon_2015} can be obtained from Eq.~(\ref{eq:gk}) as special cases.
\subsection{Second sound}
The right side (RHS) of the G-K equation represents the effect of viscosity on
the heat transport behavior.  They come from the first order term in the
expansion over $\varepsilon$.  Before looking into these terms, we show here
that the propagation of second sound can be analyzed without these terms.
Replacing the RHS with zero, combining with Eq.~(\ref{eq:econsv}), we arrive at
\begin{equation}
\frac{\partial^2 T}{\partial t^2} +\frac{1}{\tau_R} \frac{\partial T}{\partial t} - v^2_{ss} \nabla^2 T = 0,
\end{equation}
where we have defined the second sound velocity
\begin{equation}
v^2_{ss}=\frac{\kappa_0}{C\tau_R}=\alpha \gamma v_g^2.
\end{equation}
This is the wave equation describing propagation of second sound with velocity
$v_{ss}$ and damping coefficient $\tau_R^{-1}$.  For 3D materials with the
Debye model, the second sound velocity $v_{ss}=v_g/\sqrt{3}$, similar model for
2D material gives $v_{ss}=v_g/\sqrt{2}$.
Here, the presence of quadratic
dispersion makes $v_{ss}$ temperature dependent, inherited from the different
temperature dependence of $C_{L}$ and $C_{N}$.

\subsection{Poiseuille flow}
We now include the RHS of the G-K equation, and consider a nano-ribbon with
length $L$ ($0\le x \le L$) and width $w$ ($0\le y \le w$) [Fig.~\ref{fig:flow}
(a)].  A temperature difference is applied along the ribbon ($x$ direction)
[Fig.~\ref{fig:flow} (b)]. At steady state, ignoring $\Vq/\tau_R$,
Eq.~(\ref{eq:gk}) reduces to a one-dimension form $\frac{\partial^2 q}{\partial
y^2} = A $.  This gives rise to a parabolic heat flux distribution
perpendicular to the flow $q(y) = \frac{A}{2} y (y-w)$, with $ A = (\partial
T/\partial x)\kappa_0/\tau_R\eta$, if we assume a non-slip boundary condition
$q(0)=q(w)=0$. By integration over $y$, the heat current is obtained
\begin{equation}
I =\int_0^w q(y) dy=-\frac{1}{12}A w^3.
\end{equation}
The negative sign means heat flows opposite to the temperature gradient. The heat current scaling as $w^3$ is a
signature of the Poiseuille flow.
For diffusive phonon transport, the heat current scales linearly with the
ribbon width $I\propto w$, while for ballistic transport, the heat current
can not go higher than linear scaling with the width. Thus, the cubic (super-linear) 
dependence of $I$ on $w$ can in
principle be used as a signature of  the Poiseuille flow.  The Poiseuille flow
in graphene ribbons has been studied numerically by solving the Boltzmann
equation directly in
Refs.~\onlinecite{lee_hydrodynamic_2015,cepellotti_phonon_2015}.  Similar
behavior is also predicted for electronic transport in graphene
nano-ribbons\cite{levitov_electron_2016}.  Length and width dependent thermal
conductivity in suspended single layer graphene has been reported
experimentally\cite{bae_ballistic_2013,Xu2014}. Thus, experimental confirmation of Poiseuille flow
in graphene is already within reach.

\begin{figure*}
\includegraphics[scale=1]{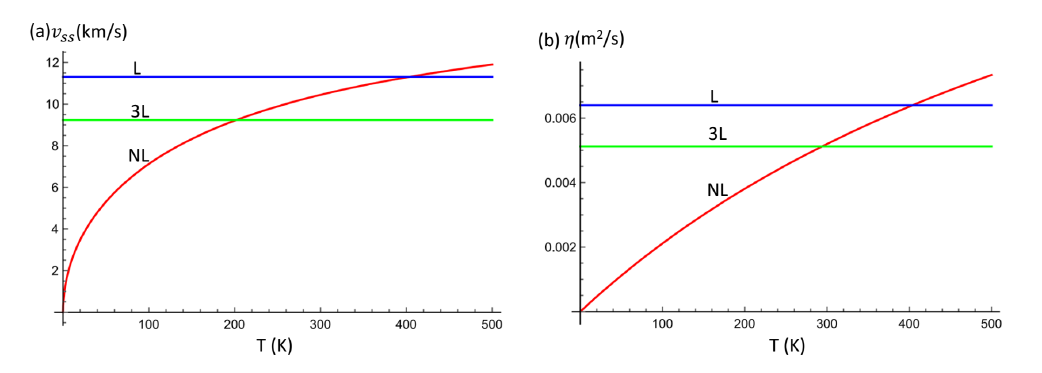}
\caption{(a) The second sound velocity $v_{ss}$ as a function of temperature (T) for three different situations. (b) The dependence of the first viscosity coefficient $\eta$ on temperature (T). Here, we take the $\tau_N=10^{-10}$ s. NL stands for the 2D material with one quadratic ZA mode and two degenerate linear acoustic modes (longitudinal and transverse). L represents the 2D material with three degenerate linear acoustic phonons modes (2D Debye model). 3L means 3D material with three linear degenerate acoustic phonon modes (3D Debye model). }
\label{fig:withT}
\end{figure*}

\section{Heat current vortices and negative nonlocal conductance}
\label{sec:nonlocal}
One important consequence of Eq.~(\ref{eq:gk}) is nonlocal heat transport and
formation of heat current vortices when there is a heat current source injecting
into the 2D materials. As far as we know, this has not been considered before. To show this, we consider steady state transport in a
setup sketched in Fig.~\ref{fig:flow} (c).
Heat current source and drain are attached to a graphene nano-ribbon. The
pattern of current flow at steady state can be obtained from the solution of
simplified version of Eq.~(\ref{eq:gk}). At steady state, according to
Eq.~(\ref{eq:econsv}), we have $\partial \vec{q}/\partial
t=\partial E/\partial t=0$. The resulting equation has the form
\begin{equation}
\eta \nabla^2 \vec{q} -\tau_R^{-1} \vec{q}= C_{L}v_g^2 \nabla T.
\label{eq:ssgk}
\end{equation}
It shares the same form as the electronic case in
Ref.~\onlinecite{levitov_electron_2016}.  The 1st term on the left hand side
(LHS) is the hydrodynamic term due to viscosity. It is the origin of the
nonlocal transport behavior. When $\tau_R^{-1}$ is negligible, we get pure
hydrodynamic viscous flow. One the other hand, when $\eta$ is negligible, we
recover the normal diffusive heat transport governed by Fourier law.  
Thus, equation~\ref{eq:ssgk} applies to steady state diffusive and hydrodynamic transport. Actually, it is suggestive to
define a dimensionless parameter
\begin{equation}
\epsilon = \frac{w^2 }{\eta\tau_R} = \frac{w^2}{\beta v_g^2 \tau_N \tau_R}
\end{equation}
to characterize the relative contribution of diffusive and hydrodynamic transport.

As an example, we have plotted typical heat current flow patterns (lines) and the resulting
temperature distribution (color) with $\epsilon=1$ and $300$ in Fig.~\ref{fig:flow} (c) and (d), respectively. Here, a
flow of heat current from a point source $I(x)=I \delta(x)$ is injected into the ribbon and
collected at the opposite side. The non-slip boundary condition is used to solve Eq.~(\ref{eq:ssgk}). The heat current
flow within the ribbon can be obtained by solving Eq.~(\ref{eq:ssgk}) with the
help of the streaming function. In Fig.~\ref{fig:flow} (c), hydrodynamic transport is dominant. The formation of vortices besides both sides
of the direct source-to-drain flow is a characteristic feature of the viscous flow.
Another prominent feature is the `separation' of temperature gradient and heat flow.
Even negative thermal resistance can be observed, where the heat current flows
from the low to the high temperature regime.
In Fig.~\ref{fig:flow} (d), the diffusive Fourier transport becomes dominant.
Heat current vortices and negative resistance are absent.
Thus, we can study the transition from hydrodynamic to Fourier transport by changing the magnitude of $\epsilon$.

We now give an order-of-magnitude analysis using parameters of graphene.  We obtain the phonon dispersion relation of graphene using density
function theory based calculations \footnote{For the density functional theory calculation, we use the Vienna Ab-initio
Simulation Package and the generalized gradient approximation for the
exchange-correlation functional. The parameters are the same as
Ref.~\onlinecite{Ge2016}.}.  Fitting the dispersion relation results in
$v_g=1.6\times10^{4}$ m/s for the linear modes, $a=5.5\times10^{-7}$ m$^2$/s for
the quadratic mode.  The specific heat capacity of them is given by
$C_{L}=\frac{6Zeta(3)}{\pi}\frac{k_B^3 T^2}{(\hbar v_g)^2}\approx 2.14\times
10^{-9} T^2$ Jm$^{-2}$K$^{-1}$ and $C_{N}=\frac{\pi}{12}\frac{k_B^2 T}{\hbar
a}\approx 8.63\times 10^{-7} T$ Jm$^{-2}$K$^{-1}$, respectively.
To estimate the transport coefficients and the dimensionless factor $\epsilon$, we use $\tau_N=10^{-10}$ s,
$\tau_R=10^{-7}$ s \cite{lee_hydrodynamic_2015}.
We get $\beta \approx 0.08$ at $T=100$ K, smaller than value obtained from the 2D Debye model $\beta=1/4$.
The different temperature dependence of $C_L$, $C_N$ and $E_L$,  $E_N$ gives rise to temperature dependent $v_{ss}$ and $\eta$. This is contrast to 2D or 3D Debye model. The comparison of different situations is shown in Fig.~\ref{fig:withT} (a) and (b) for $v_{ss}$ and $\eta$, respectively. We find that, the presence of quadratic ZA mode reduces $v_{ss}$ and $\eta$ in the relevant temperature $\sim 100$ K. Still, we get $\eta\approx 0.002$ m$^2$/s at $T=100$ K, which is orders of magnitude larger than that of water. 
We also get the dimensionless parameter  $\epsilon \approx 5 \times 10^{9} w^2 $ m$^{-2}$.  Thus, phonon hydrodynamic transport
can be realized in  high quality graphene nano-ribbons of micrometer scale. The plot in Fig.~\ref{fig:flow} (c) with
$\epsilon=1$ corresponds to sample length of $\sim 15 $ $\mu$m.

\section{Conclusions}
\label{sec:conclusion}
In summary, we have derived a 2D version of the G-K equation describing
hydrodynamic phonon heat transport.  We take into account the out of plane
quadratic phonon dispersion of the ZA mode, normally present in 2D materials.
Its effect on the hydrodynamic transport is analyzed. The derived equation
serves as a starting point for investigating nonlocal hydrodynamic phonon
transport behavior in 2D materials. It shares similar form as the Navier-Stokes
equation that has been used to study electron hydrodynamic transport in
graphene. Many interesting transport behaviors, including nonlocal negative
resistance, higher-than-ballistic transport, predicted for electrons can be
studied for phonons. Moreover, a large overlap of the parameter regime between
electron and phonon hydrodynamic transport in graphene makes it promising to
study the effect of their mutual interaction on the thermal transport behavior
of the two kinds of fluids. 

During the submission of present work, several new studies came out\cite{zhang_implicit_2019,PhysRevB.97.035421,PhysRevB.97.094309}, which are all based on numerical calculations and complementary to present analytical study.

\acknowledgements
The authors thank Nuo Yang, Jin-Hua Gao for
discussions. We acknowledge financial support from National Natural Science
Foundation of China (Grant No. 21873033).

\appendix
\onecolumngrid
\setcounter{figure}{0}
\section{Derivation of the 2D G-K equation}
\label{sec:appa}
Here, we give the derivation of Eq.~(\ref{eq:gk}) in the main text. We use the recent developed multiscale
expansion technique\cite{guo_phonon_2015}. The expansion is over both space and time as
follows:
  \begin{align}
  \frac{\partial{}}{\partial{t}}=&\varepsilon\frac{\partial}{\partial{t_1}}
  +\varepsilon^2\frac{\partial}{\partial{t_2}},\\
  \frac{\partial{}}{\partial{x_{i}}}=&\varepsilon\frac{\partial}{\partial{x_{1i}}},
  \end{align}
where $\varepsilon$ is a small parameter, defined as
\begin{equation}
\varepsilon = \frac{\tau_N}{\tau_R}.
\end{equation}
That is to say, we consider the situation where the scattering rates of N-process is much larger than that of R-process, preferential for hydrodynamic phonon transport. We expand the phonon distribution function as follows
  \begin{equation}
  f=f_0+\varepsilon f_1+\varepsilon^2 f_2+\cdots,
  \end{equation}
with $f_0$, $f_1$, $f_2$ the 0th, 1st and 2nd order terms of phonon distribution function. The macroscopic variables can be expressed by the sum of approximate components
  \begin{align}
  E&=E_0+\varepsilon E_1+\varepsilon^2 E_2+\cdots ,\quad
  E_n=\sum_s \int\hbar\omega_{\sk} f_{n,\sk}\frac{d\VK}{(2\pi)^2},\\
  \Vq&=\vec{q}_0+\varepsilon \vec{q}_1+\varepsilon^2 \vec{q}_2+\cdots ,\quad
  \vec{q}_n=\sum_s \int\vec{v}_{\sk}\hbar\omega_{\sk} f_{n,\sk}\frac{d\VK}{(2\pi)^2},\label{eq:eqq}\\
 \vec{\kappa}&=\vec{\kappa}_0+\varepsilon \vec{\kappa}_1+\varepsilon^2 \vec{\kappa}_2+\cdots , \quad \vec{\kappa}_n=\tau_R \sum_s \int\hbar\omega_{\sk}\vec{v}_{\sk}\vec{v}_{\sk}\frac{\partial f_{n,\sk}}{\partial  T}\frac{d\VK}{(2\pi)^2}.
  \end{align}
Substituting all of the equations above into the Peierls-Boltzmann equation, we obtain the 0th and 1st order terms of phonon distribution function:
  \begin{align}
  f_0&=f_N^{eq},\\
  f_1&=f_R^{eq}-f_N^{eq}-\tau_N\left(\frac{\partial{f_N^{eq}}}{\partial t_1}+v_{i}\frac{\partial{f_N^{eq}}}{\partial x_{1i}}\right),
  \end{align}
and different order of approximations for the balance equations
  \begin{align}
  \frac{\partial{E}}{\partial{t_1}}&+\frac{\partial{}}{\partial{x_{1i}}}q_{0i}=0,
  \label{eq:ae0}\\
  \frac{\partial{E}}{\partial{t_2}}&+\frac{\partial{}}{\partial{x_{1i}}}q_{1i}=0,
  \label{eq:ae1}\\
  \frac{\partial{q_{0i}}}{\partial{t_1}}&+\kappa_{0ij}\frac{\partial{T}}{\partial{x_{1j}}}
  =-\frac{1}{\tau_N}q_{0i}-\frac{1}{\tau_N}q_{1i},
  \label{eq:aq0}\\
  \frac{\partial{q_{0i}}}{\partial{t_2}}&+\frac{\partial{q_{1i}}}{\partial{t_1}}+
  \kappa_{1ij}\frac{\partial{T}}{\partial{x_{1j}}}=-\frac{1}{\tau_N}q_{1i}
  -\frac{1}{\tau_N}q_{2i}.
  \label{eq:aq1}
  \end{align}

\subsection{The zeroth-order result}
We calculate the macroscopic variables including the energy density, the heat flux and the thermal conductivity, taking into account the 0th order term $f_0$. We make the following approximation
\begin{equation}
f_0 = f_{N}^{eq} = f_{R}^{eq} + f_{N1}^{eq} \approx f_{R}^{eq}+\beta_B f_{R}^{eq}(f_{R}^{eq}+1) \hbar\VK\cdot\Vu,
\end{equation}
whose validity has been checked numerically\cite{lee_hydrodynamic_2015}.
\subsubsection{Energy density}
We consider the phonon energy density first.
According to energy conservation (\ref{eq:ec1}), we know that $E=E_0$,
\begin{align}
  E=\sum_s \int\hbar\omega_{\sk} f_{N,\sk}^{eq}\frac{d\VK}{(2\pi)^2}
  =\sum_s \int\hbar\omega_{\sk} f_{R,\sk}^{eq}\frac{d\VK}{(2\pi)^2}+\sum_s \int\hbar\omega_{\sk} f_{R,\sk}^{eq}\left(1+f_{R,\sk}^{eq}\right)\frac{\hbar\VK\cdot\Vu}{k_BT}\frac{d\VK}{(2\pi)^2}.
  \end{align}
The second term in the above equation vanishes because it's an odd function of  $\VK$. Thus we arrive at:
  \begin{align}
  E=\sum_s \int\hbar\omega_{\sk} f_{R,\sk}^{eq}\frac{d\VK}{(2\pi)^2} =\frac{2}{\pi}\frac{(k_BT)^3}{(\hbar v_g)^2}Zeta(3)+\frac{\pi}{24}\frac{(k_BT)^2}{\hbar a} \equiv E_L+E_N.
  \end{align}
Here, $E_L$ and $E_N$ are the energy density of linear and quadratic phonon modes, respectively. Note that we have included the factor 2 to account the degeneracy of the linear phonon modes.

\subsubsection{Heat flux}
The 0th order heat flux is calculated as
\begin{align}
  \Vq_0=\sum_s \int\vec{v}_{\sk}\hbar\omega_{\sk} f_{N,\sk}^{eq}\frac{d\VK}{(2\pi)^2}
  =\sum_s \int\vec{v}_{\sk}\hbar\omega_{\sk} f_{R,\sk}^{eq}\frac{d\VK}{(2\pi)^2}
  +\sum_s \int\vec{v}_{\sk}\hbar\omega_{\sk} f_{N1,\sk}^{eq}\frac{d\VK}{(2\pi)^2}.
  \end{align}
Since at thermal equilibrium, the heat flux is zero. Only the second term contributes
  \begin{align}
  \Vq_0=\sum_s \int\vec{v}_{\sk}\hbar\omega_{\sk} f_{N1,\sk}^{eq}\frac{d\VK}{(2\pi)^2}
  =\frac{3}{\pi}\frac{(k_BT)^3}{(\hbar v_g)^2}Zeta(3)\Vu+
  \frac{\pi}{12}\frac{(k_BT)^2}{\hbar a}\Vu
  \equiv \vec{q}_{0L}+\vec{q}_{0N}.
  \end{align}
Here, $\vec{q}_{0L}$, $\vec{q}_{0N}$ are the 0th order contribution of the
linear and quadratic modes to the heat flux.

\subsubsection{Thermal conductivity tensor}
The thermal conductivity tensor is
  \begin{align}
  \vec{\kappa}_0 &= \tau_R \sum_s \int\hbar\omega_{\sk}\vec{v}_{\sk}\vec{v}_{\sk}\frac{\partial f_{N,\sk}^{eq}}{\partial T}\frac{d\VK}{(2\pi)^2}
  =\tau_R \sum_s \int\hbar\omega_{\sk}\vec{v}_{\sk}\vec{v}_{\sk}\frac{\partial f_{R,\sk}^{eq}}{\partial T}\frac{d\VK}{(2\pi)^2}
  +\tau_R\sum_s \int\hbar\omega_{\sk}\vec{v}_{\sk}\vec{v}_{\sk}\frac{\partial f_{N1,\sk}^{eq}}{\partial T}\frac{d\VK}{(2\pi)^2}.
  \end{align}
Likewise, the second term in the above equation is an odd function of $\VK$ and thus is zero. We then have
  \begin{align}
  \vec{\kappa}_0 &=\tau_R\sum_s \int\hbar\omega_{\sk}\vec{v}_{\sk}\vec{v}_{\sk}\frac{\partial f_{R,\sk}^{eq}}{\partial T}\frac{d\VK}{(2\pi)^2}
  =\tau_R \sum_s \int\hbar\omega_{\sk}\vec{v}_{\sk}\vec{v}_{\sk}f_{R,\sk}^{eq}\left(f_{R,\sk}^{eq}+1\right)\frac{\hbar\omega_{\sk}}{k_BT^2}\frac{d\VK}{(2\pi)^2}\notag \\
  &=\frac{3}{\pi}\frac{k_B^3T^2}{\hbar^2}Zeta(3)\tau_R \VI+
  \frac{3}{\pi}\frac{k_B^3T^2}{\hbar^2}Zeta(3)\tau_R \VI
  \equiv\vec{\kappa}_{0L}+\vec{\kappa}_{0N}.
  \end{align}
We get the 0th order thermal conductivity contributed from the linear $\vec{\kappa}_{0L}$ and quadratic modes  $\vec{\kappa}_{0N}$.
It is noted that, given a constant $\tau_R$, the thermal conductivity contributed from
a quadratic mode is two times larger than that contributed from a linear mode. Both of
them are independent of the details of the dispersion.
So we can rewrite $\kappa_0$ in terms of the energy density of the linear phonon modes,
  \begin{equation}
  \vec{\kappa}_0=\frac{6Zeta(3)}{\pi}\frac{k_B^3T^2}{\hbar^2}\tau_R\VI=C_{L}v_g^2\tau_R\VI,
  \end{equation}
  where the specific heat capacity of the linear modes is:
  \[C_{L}=\frac{\partial E_L}{\partial T}=\frac{6Zeta(3)}{\pi}\frac{k_B^3}{(\hbar v_g)^2}T^2.\]

\subsection{The first-order result}
Now, we calculate the 1st order terms in the expansion
 \begin{align}
 f_1&=f_R^{eq}-f_N^{eq}-\tau_N\left(\frac{\partial{f_N^{eq}}}{\partial t_1}+v_{i}\frac{\partial{f_N^{eq}}}{\partial x_{1i}}\right).
 \end{align}
Based on the chain rule
\begin{align}
\frac{\partial{f_N^{eq}}}{\partial{t_1}}&=
\frac{\partial{f_N^{eq}}}{\partial{u_j}}\frac{\partial{u_j}}{\partial{t_1}}+
\frac{\partial{f_N^{eq}}}{\partial{T}}\frac{\partial{T}}{\partial{t_1}},\\
\frac{\partial{f_N^{eq}}}{\partial{x_{1i}}}&=
\frac{\partial{f_N^{eq}}}{\partial{u_j}}\frac{\partial{u_j}}{\partial{x_{1i}}}+
\frac{\partial{f_N^{eq}}}{\partial{T}}\frac{\partial{T}}{\partial{x_{1i}}},
\end{align}\\
and the approximation for $f_N^{eq}$,
combined with Eqs.~(\ref{eq:ae0}),(\ref{eq:ae1}), we arrive at a lengthy expression for $f_1$ :
\begin{multline}
\label{eq:f11}
f_1=-\frac{\hbar\VK\Vu}{k_BT}f_R^{eq}\left(f_R^{eq}+1\right) -\tau_Nf_R^{eq}\left( f_R^{eq}+1\right) \left\{\frac{\hbar k_i}{k_BT}
\left( -\frac{2C_{L}v_g^2}{3E_L+4E_N}\frac{\partial T}{\partial x_{1i}}+
\frac{u_i}{3E_L+4E_N}\frac{3C_{L}+4C_{N}}{C}\frac{\partial q_{0j}}{\partial x_{1j}}-\right.\right.\\
\left.\left.\frac{2}{3E_L+4E_N}\frac{q_{0i}}{\tau_N} \right)
+\left(\frac{v}{k}k_i\frac{\partial T}{\partial{x_{1i}}}-\frac{1}{C}\frac{\partial q_{0j}}{\partial x_{1j}}\right)
\left[\frac{\hbar\omega}{k_BT^2}+\frac{\hbar\omega}{k_BT^2}\left(2f_R^{eq}+1\right)\frac{\hbar\VK\Vu}{k_BT}
-\frac{\hbar\VK\Vu}{k_BT^2}\right]
+\frac{v}{k}\frac{\hbar k_ik_j}{k_BT}\frac{\partial}{\partial{x_{1i}}}\left(\frac{2q_{0j}}{3E_L+4E_N}\right)\right\}.
\end{multline}

We notice from eqs.~(\ref{eq:ae0}-\ref{eq:aq1}) that, $f_1$ depends on $\vec{q}_1$, while to obtain $\vec{q}_1$ we need $f_1$. In principle, we need to self-consistent calculations of them. Here, to arrive at analytical result, we made the truncation and set $\vec{q}_1=0$ in eqs.~(\ref{eq:ae0}-\ref{eq:aq1}).

\subsubsection{Phonon energy density}
According to the foregoing discussion, we can get :
  \begin{align}
  E&=E_0.
  \end{align}
The higher order terms of distribution function do not contribute to the energy density.

\subsubsection{Heat flux}
Using the truncation we made for $f_1$, we can check that $\Vq_1=0$ if we plug
Eq.~(\ref{eq:f11}) into Eq.~(\ref{eq:eqq})
\begin{align}
\Vq_1&=\sum_s \int\hbar\omega_{\sk}\vec{v}_{\sk}f_{1,\sk}\frac{d\VK}{(2\pi)^2}=\Vq_{1L}+\Vq_{1N}=0,
\end{align}
as required by our truncation scheme.

\subsubsection{Thermal conductivity tensor}
Finally, we consider the thermal conductivity tensor. We consider the linear
and quadratic modes separately
\begin{align}
\vec{\kappa}_1&=\tau_R \frac{\partial}{\partial T}\left(\vec{Q}_L+\vec{Q}_N\right),
\end{align}
with $\vec{Q}_L$, $\vec{Q}_N$ corresponding to the contribution from linear and quadratic modes,  respectively. Again, the odd parts do not contribute to the integral. Collecting the even parts, we divide $\vec{Q}_L$ into three parts:
\begin{align}
\vec{Q}_L=&\vec{Q}_{L,\uppercase\expandafter{\romannumeral1}}
+\vec{Q}_{L,\uppercase\expandafter{\romannumeral2}}+
\vec{Q}_{L,\uppercase\expandafter{\romannumeral3}},
\end{align}
with each part written as
\begin{align}
(Q_{L,\uppercase\expandafter{\romannumeral1}})_{mn}
&=\frac{1}{2}\tau_Nv_g^2\frac{C_{L}}{C}\frac{\partial q_{0j}}{\partial x_{1j}}\delta_{mn},\\
(Q_{L,\uppercase\expandafter{\romannumeral2}})_{mn}
&=-\frac{3}{4}v_g^2E_L\tau_N\left[\frac{\partial}{\partial x_{1i}}\left(\frac{q_{0i}}{3E_L+4E_N}\right)\delta_{mn}+
\frac{\partial}{\partial x_{1m}}\left(\frac{q_{0n}}{3E_L+4E_N}\right)+\frac{\partial}{\partial x_{1n}}\left(\frac{q_{0m}}{3E_L+4E_N}\right)\right],\\
(Q_{L,\uppercase\expandafter{\romannumeral3}})_{mn}
&=-\frac{3}{4}\frac{v_g^2\tau_N}{T}\left[\left(q_{0L}\right)_i\frac{\partial T}{\partial x_{1i}}\delta_{mn}+
\left(q_{0L}\right)_n\frac{\partial T}{\partial x_{1m}}+\left(q_{0L}\right)_m\frac{\partial T}{\partial x_{1n}}\right].
\end{align}
Similarly,
\begin{align}
\vec{Q}_N=&\vec{Q}_{N,\uppercase\expandafter{\romannumeral1}}
+\vec{Q}_{N,\uppercase\expandafter{\romannumeral2}}
+\vec{Q}_{N,\uppercase\expandafter{\romannumeral3}},
\end{align}
with
\begin{align}
(Q_{N,\uppercase\expandafter{\romannumeral1}})_{mn}
&=\frac{1}{2}\tau_Nv_g^2\frac{C_{L}}{C}\frac{\partial q_{0j}}{\partial x_{1j}}\delta_{mn},\\
(Q_{N,\uppercase\expandafter{\romannumeral2}})_{mn}
&=-\frac{3}{2}v_g^2E_L\tau_N\left[\frac{\partial}{\partial x_{1i}}\left(\frac{q_{0i}}{3E_L+4E_N}\right)\delta_{mn}+
\frac{\partial}{\partial x_{1m}}\left(\frac{q_{0n}}{3E_L+4E_N}\right)+\frac{\partial}{\partial x_{1n}}\left(\frac{q_{0m}}{3E_L+4E_N}\right)\right],\\
(Q_{N,\uppercase\expandafter{\romannumeral3}})_{mn}
&=-\frac{3}{2}\frac{v_g^2\tau_N}{T}\left[\left(q_{0L}\right)_i\frac{\partial T}{\partial x_{1i}}\delta_{mn}+
\left(q_{0L}\right)_n\frac{\partial T}{\partial x_{1m}}+\left(q_{0L}\right)_m\frac{\partial T}{\partial x_{1n}}\right].
\end{align}
Summing up the linear and nonlinear results  gives
\begin{align}
(Q_{\uppercase\expandafter{\romannumeral1}})_{mn}
&=\tau_Nv_g^2\frac{C_{L}}{C}\frac{\partial q_{0j}}{\partial x_{1j}}\delta_{mn},\\
(Q_{\uppercase\expandafter{\romannumeral2}})_{mn}
&=-\frac{9}{4}v_g^2E_L\tau_N\left[\frac{\partial}{\partial x_{1i}}\left(\frac{q_{0i}}{3E_L+4E_N}\right)\delta_{mn}+
\frac{\partial}{\partial x_{1m}}\left(\frac{q_{0n}}{3E_L+4E_N}\right)+\frac{\partial}{\partial x_{1n}}\left(\frac{q_{0m}}{3E_L+4E_N}\right)\right],\\
(Q_{\uppercase\expandafter{\romannumeral3}})_{mn}
&=-\frac{9}{4}\frac{v_g^2\tau_N}{T}\left[\left(q_{0L}\right)_i\frac{\partial T}{\partial x_{1i}}\delta_{mn}+
\left(q_{0L}\right)_n\frac{\partial T}{\partial x_{1m}}+\left(q_{0L}\right)_m\frac{\partial T}{\partial x_{1n}}\right],
\end{align}
and $\kappa_1$ is obtained from
\begin{align}
\vec{\kappa}_1=\tau_R \frac{\partial \vec{Q}}{\partial T}=\tau_R 	\left(\nabla\cdot\vec{Q}\right)\frac{1}{\nabla T}.
\end{align}

\subsection{The G-K equation}
Substituting the calculated 0th and 1st order results into Eq.~(\ref{eq:hd0}),
neglecting the nonlinear effects resulting from the product of heat flux and temperature gradient in $\kappa_1$,  we get the G-K equation
\begin{equation}
	\frac{\partial{\Vq}}{\partial{t}}+\frac{\kappa_0}{\tau_R} \nabla T+\frac{1}{\tau_R}\Vq=\eta
\big[\nabla^2\Vq+2\nabla(\nabla\cdot\Vq)\big]-
\zeta\nabla(\nabla\cdot\Vq).
\label{eq:gk2}
\end{equation}
The transport coefficients are
$\kappa_0 = \alpha C_L v_g^2 \tau_R$,
$\eta = \beta v_g^2 \tau_N$, 
$\xi = \gamma v_g^2 \tau_N$,
with $\alpha = 1$, $\beta = \frac{9}{4} \frac{E_L}{3E_L+4E_N}$, $\gamma = \frac{C_L}{C}$, respectively.
For comparison, we also consider the 2D Debye model with three degenerate acoustic phonon modes. 
The resulting equation is the same, but with the coefficients $\alpha=\frac{1}{2}$, $\beta=\frac{1}{4}$, $\gamma=\frac{1}{2}$.
These results can be compared with the 3D Debye model, resulting in  $\alpha=\frac{1}{3}$, $\beta=\frac{1}{5}$, $\gamma=\frac{1}{3}$.

\twocolumngrid
\bibliography{extra,Hydrodynamics}

\end{document}